# Transport and Diffusion Enhancement in Experimentally Realized Non-Gaussian Correlated Ratchets

*Govind Paneru[1,2], Jin Tae Park[1], and Hyuk Kyu Pak[1,2*]*

[1]Center for Soft and Living Matter, Institute for Basic Science (IBS), Ulsan 44919, Republic of Korea

[2]Department of Physics, Ulsan National Institute of Science and Technology, Ulsan 44919, Republic of Korea

**Corresponding Author**

* Email: hyuk.k.pak@gmail.com

**ABSTRACT:** Living cells are known to generate non-Gaussian active fluctuations significantly larger than thermal fluctuations owing to various active processes. Understanding the effect of these active fluctuations on various physicochemical processes, such as the transport of molecular motors, is a fundamental problem in nonequilibrium physics. Therefore, we experimentally and numerically study an active Brownian ratchet comprising a colloidal particle in an optically generated asymmetric periodic potential driven by non-Gaussian noise having finite-amplitude active bursts, each arriving at random and decaying exponentially. We find that the particle velocity is maximum for relatively sparse bursts with finite correlation time and non-Gaussian distribution. These occasional kicks, which produce Brownian yet non-Gaussian diffusion, are more efficient for transport and diffusion enhancement of the particle than the incessant kicks of active Ornstein–Uhlenbeck noise.

**TOC GRAPHICS**

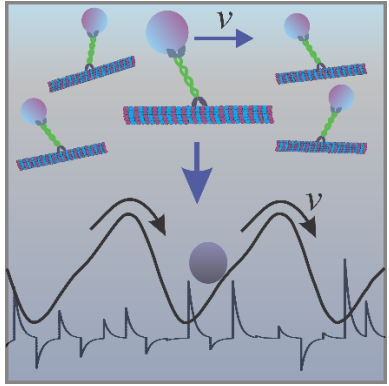

Brownian ratchets,[1-4] when driven by nonequilibrium fluctuations, can induce the directed transport of diffusive particles in asymmetric potentials. They are used as models for biological transport and nanotechnology applications, such as particle separation and the design of submicron-scale motors.[3-11] The nonequilibrium fluctuations can be generated either from nonequilibrium chemical reactions, such as ATP hydrolysis, or externally through the time-dependent perturbation of the ratchet potential. Accordingly, several Brownian ratchet models have been proposed,[4] and the conditions for their optimal operation have been studied.[4, 12-13] Some of these ratchet models have been used to explain the directed transport of the molecular motors.[11, 14] Recent single-molecule in vitro studies have revealed that chemically driven molecular motors utilize the input chemical energy for cargo transport by applying mechanical forces on their tracks.[15-17] However, the molecular motors within a living cell operate in the cell's active environment, which produces active fluctuations more significant than thermal fluctuations.[18-19] Whether these active fluctuations enhance the performance of individual molecular motors in living cells is elusive and not understood fully.

Correlated ratchets are a special class of Brownian ratchets that consider the diffusion and transport of Brownian particles in asymmetric periodic potentials in the presence of correlated active noise.[4, 20] The earlier theoretical works on the correlated ratchet mainly consider the motion of Brownian particles in the presence of Active Ornstein–Uhlenbeck (AOU) noise,[4, 21-22] which is a continuous kicking model that follows the Gaussian statistics. There are also some studies on the transport of the Brownian ratchets in the presence of non-Gaussian noises.[23-25] These studies mainly focus on the effect of the non-Gaussian noise on the velocity of the Brownian particles in the ratchet potentials. Here, we aim to study the transport and non-Gaussian diffusion of Brownian ratchets in the presence of colored Poisson noise. Moreover, there have been no systematic experimental studies carried out to explore the effect of the active fluctuations on Brownian ratchets, which may be due to the difficulties in generating the nanoscale asymmetric periodic potential and controlling the parameters of the active noise simultaneously.

Active fluctuations in living cells arise by various active processes, such as ATP hydrolysis catalyzed by molecular motors, the active transport of the molecular motors, and cytoskeleton rearrangement.[18-19, 26-28] For example, when fueled by ATP, molecular motors fire at a discrete rate following the Poisson process and generate free energy in the form of active bursts. Some of this energy is converted to mechanical work, and the remaining is dissipated internally through hidden paths.[11, 16] The recent in vivo experiments provided evidence that the resultant effect of the active processes inside the living cells is the generation of active noise having a non-Gaussian distribution with exponential side tails.[29-31] Thus, we model the intracellular active fluctuations by Exponentially Correlated Poisson (ECP) noise, having finite-amplitude active bursts, each arriving at a random interval following the Poisson counting process and decaying exponentially with the finite correlation time.[32-33] The ECP noise covers a wide range of stochastic noises. It can be non-Gaussian or Gaussian correlated noises by adjusting the noise parameters and even becomes delta-correlated Poisson or Gaussian white noise in the limiting cases. Hence, it resembles the active fluctuations of various nonequilibrium stochastic systems, such as intracellular environments,[29, 31, 34-35] the active baths of biological and synthetic swimmers,[36-43] and glassy systems.[44]

In this study, we realize a non-Gaussian correlated Brownian ratchet comprising a colloidal particle in an optically generated asymmetric periodic potential driven by ECP noise. We find that the particle velocity is maximum for a finite correlation time where each active burst decays fully before the arrival of another burst. It decreases and saturates to the value predicted by the AOU noise for a larger correlation time. In comparison to the AOU noise, where the particle is always transported along the natural direction of the ratchet, the ECP noise can drive the particle in both directions. Furthermore, our parametric studies show that the particle position distribution at any time in steady-state is always Gaussian, and the mean square displacement is proportional to the lag time, suggesting that the particle follows normal diffusion. In contrast, the distribution of the particle displacement is Gaussian or non-Gaussian, depending on the parameters of the ECP noise. We find that the particle displacement is non-Gaussian only when the noise correlation time is shorter than the average burst interval and the noise strength is greater than the thermal strength. This Brownian yet non-Gaussian diffusion[45] leads to pronounced diffusion enhancement of the particle in the periodic potential.

*Active Brownian ratchet*.– We investigated the one-dimensional motion of a Brownian particle immersed in an active bath of temperature $T$, diffusing in a spatially asymmetric periodic potential of period $L$ and barrier height $2V_0$ (see Fig. 1(a)):

$$V(x) = -V_0[\sin(2\pi x/L) + 0.25\sin(4\pi x/L)]. \tag{1}$$

The motion of the particle is governed by both thermal fluctuations $\xi(t)$ due to the solvent molecules and active fluctuations $\eta(t)$ due to the active noise source in the solution. The equation of motion for particle position $x$ is described by the overdamped Langevin equation:

$$\gamma \frac{dx}{dt} = -\frac{\partial V(x)}{\partial x} + \xi(t) + \eta(t), \tag{2}$$

where $\gamma$ is the Stokes friction coefficient, which represents the solvent viscosity. The thermal fluctuations $\xi(t)$ are modeled by Gaussian white noise of zero mean $\langle \xi(t) \rangle = 0$ and a correlation function $\langle \xi(t)\xi(s) \rangle = 2\gamma^2 D_{th}\delta(t-s)$. Here $D_{th} = k_B T/\gamma$ is the free diffusion coefficient of the particle in the thermal bath alone. The characteristic thermal relaxation time of the particle in the potential well is given by[4], $\tau_r \approx 0.08(\gamma L^2/V_0)$.

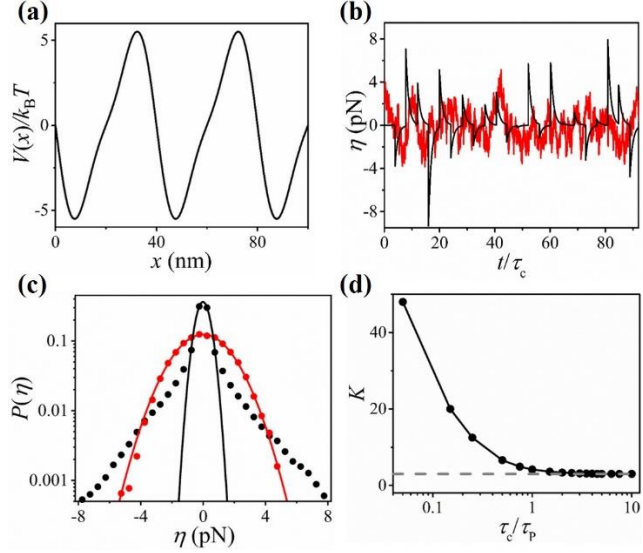

**Figure 1.** Schematic of the periodic potential and plot of the active noise: (a) Plot of the asymmetric periodic potential $V(x)$ in Eq. (1) as a function of position $x$ with $L = 40$ nm and $V_0 = 5k_B T$. As the lhs of the each potential well is steeper than the rhs, the natural ratchet direction is along the positive $x$ direction. (b) Traces of the active AOU noise (red) and ECP noise (black) of the same strength $f_{act} = \sqrt{F/(1+\lambda)} = 1.5$ pN and correlation time $\tau_c/\Delta t = 100$, where $\Delta t$ is the noise input interval, with Poisson parameter $\lambda = 0$ (red, AOU) and $\lambda = 400$ (black, ECP). (c) Probability distribution function of the active noises in panel (b). The like color solid curves are the Gaussian fit to the data. (d) Kurtosis of the active noise $K(t)$ as a function of $\tau_c/\tau_P$, showing the degree of non-Gaussianity of the ECP noise. The horizontal dashed line corresponds to a Gaussian distribution $K = 3$.

The active fluctuations $\eta(t)$ are modeled by ECP noise having finite-amplitude random kicks with correlation time $\tau_c$ arriving at an average Poisson interval $\tau_P$ following the Poisson process. $\eta(t)$ is generated numerically from Poisson white noise,[46] $\xi_{PN}(t) = \sum_{i=1}^{N(t)} y_i \delta(t-t_i)$, where $t_i$ are the arrival time of a Poisson counting process $N(t)$ with mean arrival time $\tau_P$ and $y_i$ are Gaussian random variables with variance $F$, using the recursion relation,[33, 47]

$$\eta_n = \eta_0 \exp(-n/\tau_c) + \sqrt{1-\exp(-2/\tau_c)} \sum_{i=1}^{n} (\xi_{PN})_i \exp(-(n-i)/\tau_c). \tag{3}$$

The stochastic equation governing Eq. (3) is equivalent to the active Orenstein-Uhlenbeck process $\tau_c d\eta/dt = -\eta(t) + \sqrt{2}\xi_{PN}(t)$, where $\xi_{PN}(t)$ is white Poisson noise with zero mean and $F/(1+\lambda)$ variance. Here, $\lambda$ is the Poisson parameter that gives the average number of waiting events between successive kicks. Therefore, the average time between two consecutive kicks is given by $\tau_P = \lambda \Delta t$, with a noise input interval $\Delta t$. Equation (3) generates a sequence of ECP noise of zero mean, $<\eta(t)> = 0$, with the correlation function:

$$\langle \eta(t)\eta(s) \rangle = \frac{F}{\lambda+1} \exp(-|t-s|/\tau_c). \tag{4}$$

Here, $f_{act} \equiv \sqrt{F/(1+\lambda)}$ characterizes the strength of the ECP noise. When $\lambda = 0$ (which corresponds to $\tau_P = 0$), our noise model follows AOU noise with vanishing mean $\langle \eta(t) \rangle = 0$ and correlation function $\langle \eta(t)\eta(s) \rangle_{OU} = (\gamma^2 D_{act}/\tau_c) \exp(-|t-s|/\tau_c)$. $F$ can then be expressed in terms of the active diffusion coefficient $D_{act}$ and noise correlation time $\tau_c$ as $F = \gamma^2 D_{act}/\tau_c$. For direct comparison, the time traces of the ECP and AOU noise of the same strength and correlation time are depicted in Fig. 1(b). Here, in contrast to the continuous kicks of the AOU noise, the ECP noise produces discrete kicks of random amplitudes. Note that the current noise generation approach is similar to our previous study[32] except that $\tau_c$ and $\tau_P$ can be independently varied even for the limiting case of $\tau_c = 0$ or $\tau_P = 0$.

For $\tau_c \lesssim \tau_P$, each kick decays completely on an average before the arrival of another burst, and the ECP noise follows a non-Gaussian distribution with exponential tails (black solid circles in Fig. 1(c)). The degree of non-Gaussianity of the probability distribution function (PDF) of the noise can be quantified by the kurtosis $K(t) = <[\eta(t)-<\eta(t)>]^4>/<[\eta(t)-<\eta(t)>]^2>^2$. For a Gaussian distribution, $K=3$. Figure 1(d) shows $K$ decreases with the increase in $\tau_c$ and saturates to 3 for $\tau_c \gtrsim 5\tau_P$ where many active bursts arrives before each burst decays completely.[32]

*Generation of periodic potentials.–* The asymmetric periodic potential following Eq. (1) was generated experimentally using our recently developed optical feedback trap technique.[48-49] The schematic of the optical feedback trap setup is depicted in Fig. S1. This is similar to our previous setup,[33, 50-55] except for the usage of a faster quadrant photodiode (QPD, Thorlabs PDQ30C) for position detection. Here, a 2 $\mu$m diameter polystyrene sphere suspended in deionized water was trapped in a harmonic potential, $V_{op}(x,t) = (k/2)(x-x_c(t))^2$, generated by the optical tweezers, where $x_c$ is the center of the trap and $k$ is its stiffness. The QPD measures the particle position $x$ and sends the information to a field-programmable gate array (FPGA) board at a sampling time $\Delta t \approx 10$ $\mu$s using a custom-written LabVIEW FPGA program. The FPGA board computes the feedback force $f_{fd} = -\partial_x V(x)$ necessary to generate the periodic potential in Eq. (1). Numerically generated active noise $\eta(t)$ in Eq. (3) is added to the feedback force $f_{fd}$ to impose active force on the particle (the noise input interval is thus essentially the same as the sampling time $\Delta t \approx \tau_P/\lambda$). The resultant feedback force $f_{fd}(t) + \eta(t)$ is applied to the particle in the form of an optical force by a practically instantaneous shift of the trap center to $x_c(t) = x(t) - (1/k)\partial_x V(x) + \eta(t)/k$. To realize this, the FPGA updates the tuning voltage (equivalent to the resultant force) that is applied to the acousto-optic deflector (AOD) through a radiofrequency (RF) synthesizer driver for steering the laser beam center by $x_c(t)$. On repeatedly performing this procedure, the particle experiences an asymmetric potential $V(x)$ and active noise $\eta(t)$ in time.

*Experimental results.–* The number of periods of the asymmetric periodic potential to which the particle gets transported is limited by the linear working range of the QPD, for which the trap stiffness of the optical tweezers is constant. This range is approximately ±300 nm for the current setup. Therefore, starting from the fixed initial location of the trap center $x_c(0)$ and random particle position $x(0) - x_c(0)$, each experimental run records the particle trajectory for a duration $t$ during which the particle is transported by ±300 nm. The particle velocity is then measured using the relation $v = (x(t) - x(0))/t$. We fixed the barrier height of the periodic potential as $2V_0 \approx 10 k_B T$ (which is of the order of the free energy barrier for the molecular motors)[4, 11, 56-57]. The period of the asymmetric potential is selected as $L \approx 40$ nm, such that the time scale for the particle to diffuse one period of the potential $\tau_D \approx L^2/D_{th} \approx 7$ ms is considerably larger than the sampling time $\Delta t \approx 10$ μs (in addition, $\Delta t$ is smaller than $\tau_r \approx 110$ μs). The spatial period of $L \approx 40$ nm is comparable to the typical step size of Myosin V and VI motors walking along F-actin[26, 58].

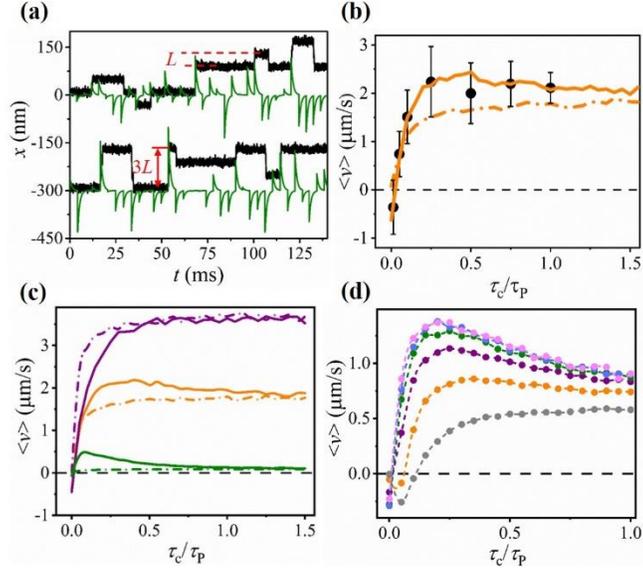

**Figure 2.** Measurement of trajectories of the particle position and velocity: (a, b experimental results): (a) Two trajectories of the particle position (black) for a $2\ \mu\mathrm{m}$ diameter particle in asymmetric potential of period $L = 40$ nm and barrier height $2V_0 = 10k_BT$ in the presence of ECP noise of strength $f_{act} \approx 1.05$ pN (greater than thermal strength $f_{th} \approx 0.8$ pN) with the Poisson interval normalized by the thermal relaxation time $\tau_P/\tau_r \approx 33.3$ and normalized correlation time $\tau_c/\tau_P \approx 0.25$. The olive curves are the corresponding active noises in time with scaled amplitude of $f_{act}/k \approx 35$ nm. For better visualization, the lower trajectory is shifted below by 300 nm. (b) Plot of the average velocity $<v>$ as a function of $\tau_c/\tau_P$ for ECP noise of the same noise strength and Poisson interval of panel (a). The error bars denote the standard error of the mean. The orange solid (orange dash dot) curve is obtained through the numerical simulation of Eq. (2) for ECP (AOU) noise under the same experimental conditions. To plot the AOU noise results, $\tau_c$ was normalized with $\tau_P$ of the corresponding ECP noise of the same strength. (c, d numerical results): (c) Plot of $<v>$ as a function of $\tau_c/\tau_P$ for ECP noise of fixed $\tau_P/\tau_r = 75$ and noise strength $f_{act} \approx 0.5$ pN (olive solid curve), 1 pN (orange solid curve), and 1.25 pN (purple solid curve). The like color dash dot curves are the plot of $<v>$ in the presence of AOU noise as a function of $\tau_c$ normalized with $\tau_P$ of the corresponding ECP noise of the same strength. (d) Plot of $<v>$ as a function of $\tau_c/\tau_P$ for ECP noise of the strength equal to the thermal strength $f_{act} \sim f_{th} \approx 0.8$ pN with Poisson interval $\tau_P/\tau_r = 6.25$ (gray), 12.5 (orange), 25 (purple), 50 (olive), 75 (blue), and 125 (magenta).

The sufficient condition for obtaining a nonvanishing mean velocity in the correlation ratchet involves energy input with finite correlation time.[22] The energy input effectively lowers the barrier height, and the particle is driven in the active force direction during the correlation time. However, because of the asymmetry of the periodic potential, particle transport along the positive *x*-direction is more favorable, which is the natural direction of the ratchet.

Figure 2(a) compares the two trajectories of a particle diffusing in the asymmetric periodic potential with the ECP noise trajectories of the strength $f_{act} \approx 1.05$ pN, the Poisson interval normalized with the relaxation time $\tau_P/\tau_r \approx 33.3$, and the correlation time normalized with the Poisson interval $\tau_c/\tau_P \approx 0.25$. One can see the barrier crossing of the particle is synchronized with the noise arrival time: The particle diffusing at the bottom of a potential well can cross the potential barrier and move to the neighboring wells when acted upon by an active burst with an amplitude comparable to the thermal strength $f_{th} = \sqrt{\gamma k_B T/\tau_r}$, i.e., $\eta(t) \sim f_{th} \approx 0.8$ pN. For larger amplitude active bursts, the particle may even cross several potential periods, which is evident from some large step jumps (see the lower trajectory in Fig. 2(a)).

Figure 2(b) depicts the average particle velocity $<v>$ as a function of $\tau_c/\tau_P$ for fixed values of $f_{act} \approx 1.05$ pN and $\tau_P/\tau_r \approx 33.3$. The average velocity was obtained by averaging more than 900 trajectories (the typical velocity distribution is shown in Fig. S2). The average velocity increases with $\tau_c$ and reaches a maximum near $\tau_c/\tau_P \approx 0.25$, for which the ECP noise still displays highly non-Gaussian behavior (see Fig. 1(d)). In contrast, for a shorter correlation time ($\tau_c = 50$ $\mu$s in Fig. 2(b)), the average velocity assumes a negative value, which indicates that the particle moves in the direction opposite to the natural direction of the ratchet.

Our experimental findings agree well with the numerical results obtained by solving Eq. (2) with Euler's method under the same experimental conditions. We also compared our results with the AOU noise of the same strength. Interestingly, we find that the particle velocity with ECP noise is greater than that with the AOU noise. This implies that the discrete kicks of the ECP noise (see Fig. 1(b)) can transport the particle with a higher velocity than the continuous kicks of the AOU noise. We will explore these findings in detail in the following numerical studies.

*Numerical results for long time limits.–* Due to the limitations in the working region of the QPD, our experimental method can only explore the intermediate time behavior of particle diffusion. To study the long-time behavior, we solved Eq. (2) numerically with $\Delta t \approx \tau_D/1000$. Figure 2(c) displays the plot of the average velocity of the particle in steady-state $<v> = \langle \lim_{t \to \infty}(x(t) - x(0))/t \rangle$ as a function of $\tau_c/\tau_P$ for $\tau_P \gg \tau_r$ and different values of active force strength $f_{act}$. Each trajectory was recorded for $t \approx 10{,}000$ $\tau_r$, and the average was taken over 1,000 trajectories. We find that, for $f_{act}$ comparable to $f_{th} \approx 0.8$ pN (olive and orange solid curves in Fig. 2(c)), $<v>$ increases with $\tau_c$ and becomes maximum at a finite value of $\tau_c < \tau_P$ which corresponds to the non-Gaussian regime of the ECP noise. Beyond this value, $<v>$ saturates to the value predicted by the AOU noise. Similar results were obtained for a $0.5$ $\mu$m diameter particle moving in the asymmetric potential of period $L = 8$ nm (equal to the typical step size of kinesin and dynein

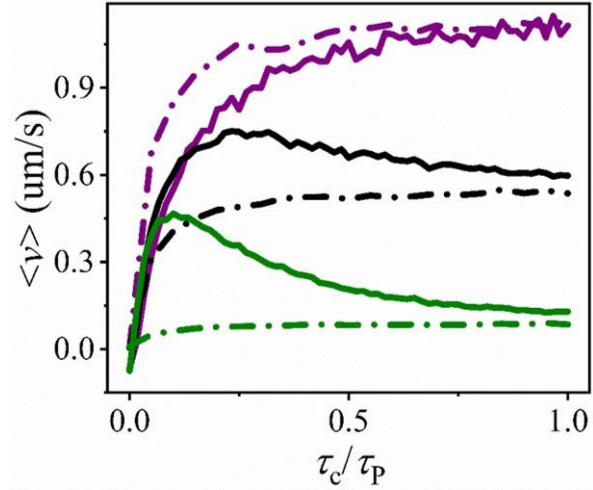

**Figure 3.** (Numerical result) Particle velocities for various barrier heights. The solid curves are the plot of $<v>$ as a function of $\tau_c/\tau_P$ for a $2\ \mu m$ diameter particle in asymmetric potential of period $L = 40$ nm in the presence of ECP noise of fixed $f_{act} \approx 0.5$ pN and $\tau_P/\tau_r = 75$ with different barrier height of $2V_0 = 5k_BT$ (purple), $7k_BT$ (black), and $10k_BT$ (olive), respectively. The like color dash dot curves are the plot of $<v>$ in presence of AOU noise as a function of $\tau_c$ normalized with $\tau_P$ of the corresponding ECP noise of the same strength.

motors[58]) and barrier height $2V_0 = 10k_BT$, as shown in Fig. S3. However, since the thermal strength for this case is $f_{th} \approx 4$ pN, the particle velocity below $f_{act} \leq 1$ pN was nominal.

For larger strength of the active force $f_{act} > f_{th}$, the average velocity increases with $\tau_c$ and saturates beyond $\tau_c/\tau_P \gtrsim 1$, where the particle experiences multiple kicks simultaneously. Note that for $f_{act} \gg f_{th}$, particle velocity with ECP noise can be smaller than the continuous kicking of the AOU noise of the same strength (purple curves in Fig. 2(c)). However, producing such incessant kicks of enormous strength in living cells would require a continuous firing of the molecular motors. Moreover, recent *in vivo* studies suggest the strength of active fluctuations in living cells is less than one piconewton,[29, 35, 59] i.e. the condition $f_{act} \gg f_{th}$ may not be feasible in living cells.

Figure 2(d) shows the dependence of the average velocity of the particle on $\tau_P \gg \tau_r$ for fixed value of $f_{act} \approx f_{th}$. We find that for $\tau_c \gtrsim 0.6\tau_r$, the average velocity increases with the increase in $\tau_P$ and becomes saturated for $\tau_P \gtrsim 5\tau_D$. On the other hand, for a shorter correlation time ($\tau_c \lesssim 0.6\tau_r$), the average velocity is negative. This agrees with the experimental result in Fig. 2(b), predicting that the particle velocity is negative only when $\tau_c < \tau_r$. This also includes the $\tau_c = 0$ limit. The particle velocity in the presence of AOU noise is always positive. In contrast to the previous study that demonstrated current inversion in symmetric Poisson white noise ($\tau_c = 0$),[23] we show it can occur for finite correlation time too, which is controlled by the active noise parameters and thermal relaxation time of the particle.

Next, we studied the transport properties of the ratchet for the active noise strength $f_{act} \approx 0.5$ pN comparable to the active fluctuations inside the living cells[29, 35, 59] for three different barrier heights (Fig. 3), $2V_0 = 5k_BT$ (purple), $7k_BT$ (black), and $10k_BT$ (olive). For smaller barrier height $2V_0 = 5k_BT$, the particle velocity in the presence of ECP noise increases with $\tau_c$ and saturates to the value predicted by AOU noise (purple curves). For larger barrier heights (black and olive solid curves in Fig. 3), the particle velocity with ECP noise is maximum, higher than corresponding AOU noise, at a finite value of $\tau_c$ where each active burst decays fully before another one arrives.

The observed maximum velocity at a finite value of $\tau_c$ in the presence of the ECP noise of relatively smaller strength ($f_{act} \sim f_{th}$) can be explained as follows. The ECP noise generates active bursts arriving at random with a mean interval $\tau_P$ and decaying with a correlation time $\tau_c$. Thus, each active burst effectively lowers the barrier height of the periodic potential for the duration $\tau_c$. At fixed noise strength $f_{act}$, the active burst strength decreases with an increase in $\tau_c$ (see Fig. S4). For $\tau_c \ll \tau_r$ (gray curve in Fig. S4), the ECP noise is equivalent to the white Poisson noise where each active burst decays immediately. The ratchet dynamics are similar to the "flashing ratchet" where particle diffuses freely during active kicks. As a result, the particle may escape the left barrier top more frequently, generating a negative average velocity.[23] For $\tau_r < \tau_c < \tau_P$, the ECP noise has relatively large-amplitude active bursts (blue curve in Fig. S4), which can drive the particle for a duration $\tau_c$ along its direction. Such large-amplitude highly correlated active bursts can generate "power strokes" that can transport the particle several periods of the periodic potential. However, due to the asymmetry, the positive kicks transport the particle more than the similar strength negative kicks, generating positive velocity. For $\tau_c > \tau_P > \tau_r$, the ECP noise is highly correlated, but active burst amplitudes are significantly small (orange curve in Fig. S4). Thus, the maximum velocity occurs at a finite value of $\tau_c$ where active burst strengths are large enough to drive the particle in the neighboring potential wells (blue curve in Fig. S4).

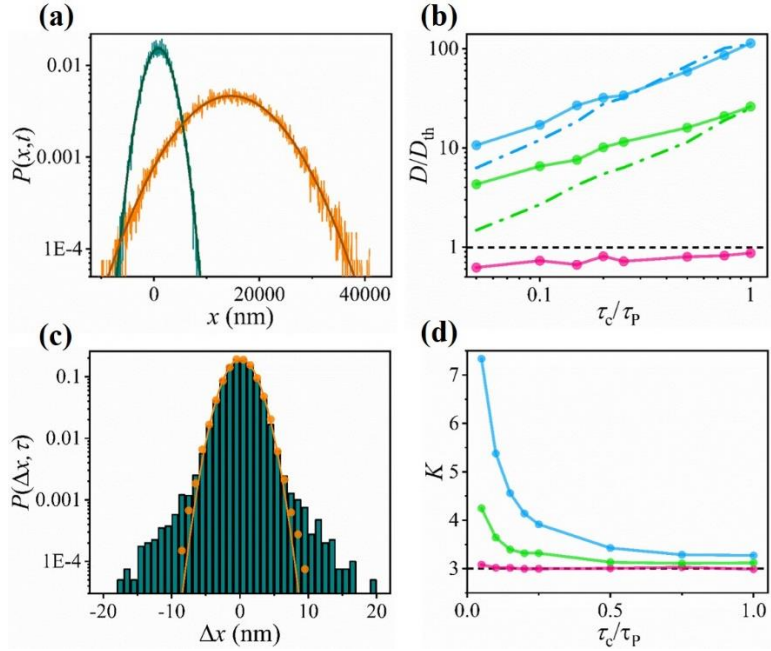

**Figure 4.** (Numerical result) Steady state probability distribution functions and effective diffusion coefficient for 2 $\mu$m diameter particle in asymmetric potential of period $L = 40$ nm and barrier height $2V_0 \approx 10 k_B T$: (a) PDF $P(x,t)$ of particle position $x$ at time $t$ for ECP noise of fixed strength $f_{act} = 3$ pN and Poisson interval $\tau_P/\tau_r = 25$ with $\tau_c/\tau_P = 0.05$ (olive) and 1 (orange). The solid curves are the Gaussian fit to the data. The PDFs are calculated in the long time limit $t \approx 10,000 \tau_r$, for which the particle diffusion is normal, $\sigma_x^2(t) = 2Dt$. The distributions are constructed from the end data points of 40,000 trajectories. (b) Plot of the effective diffusion coefficient normalized with the free thermal diffusion coefficient $D/D_{th}$ as a function of $\tau_c/\tau_P$ for ECP noise of fixed $\tau_P/\tau_r = 25$ and $f_{act} = 1$ (magenta), 2 (green), and 3 (blue) pN. The dashed-dotted curves are the plots for AOU noise of the same strength as the ECP noise data of like color. (c) PDF $P(\Delta x, \tau)$ for particle displacement $\Delta x = x(t+\tau) - x(t)$ obtained from the like-colored data in panel (a). The solid curve is the Gaussian fit to the orange data. (d) Kurtosis of $P(\Delta x, \tau)$ vs $\tau_c/\tau_P$ for ECP noise of fixed $\tau_P/\tau_r = 25$ and $f_{act} = 1$ (magenta), 2 (green), and 3 (blue) pN.

*ECP noise produces Brownian yet non-Gaussian diffusion.–* The ECP noise can lead to the Brownian yet non-Gaussian diffusion, where the mean square displacement of the particle increases linearly with time, and the PDF of the position displacement is non-Gaussian.[45, 60] To this end, we measured the PDF $P(x,t)$ of particle position $x$ in the long time limit $t \approx 10,000\tau_r$ in the presence of ECP noise of fixed strength $f_{act} = 3$ pN and Poisson interval $\tau_P/\tau_r = 25$ (Fig. 4(a)). The $P(x,t)$ always follows Gaussian distribution with kurtosis $K(t) \approx 3$ following the central limit theorem. The variance of the distribution $\sigma_x^2(t)$ increases with $\tau_c$ (orange curve in Fig. 4(a) is wider than the olive). The comblike structures detected in the PDF are due to the rapid relaxation of the particle towards the adjacent potential minima of the periodic potential.

In the long time limit $t \gg \tau_r$, $\sigma_x^2(t)$ increases linearly with time. Based on this, we measured the effective diffusion coefficient $D = \lim_{t\to\infty} \sigma_x^2(t)/2t$ of the particle in the presence of active noise. For $f_{act} \lesssim f_{th}$, although ECP noise can induce directed motion of the particle with finite velocity, the effective diffusion coefficient is less than the free thermal diffusion coefficient, $D/D_{th} < 1$ (Fig. 4(b)). For $f_{act} \gtrsim f_{th}$, the effective diffusion coefficient increases with $\tau_c$, leading to enhanced diffusion. Interestingly, diffusion enhancement for a relatively shorter correlation time is larger than that with AOU noise (see the dashed-dotted curves in Fig. 4(b)).

We also measured the PDF $P(\Delta x, \tau)$ of particle displacement $\Delta x = x(t+\tau) - x(t)$ for $\tau = \Delta t$. We found that when the active noise strength is less than the thermal strength $f_{act} \lesssim f_{th}$, $P(\Delta x, \tau)$ exhibits Gaussian behavior ($K \approx 3$) regardless of the values of $\tau_c$ and $\tau_P$. In contrast, for $f_{act} > f_{th}$ and $\tau_c < \tau_P$, where ECP noise is non-Gaussian, $P(\Delta x, \tau)$ adopts non-Gaussian behavior, exhibiting a Gaussian shape at the center and exponentially decaying side-tails (Fig. 4(c)). The degree of non-Gaussianity decreases with the increase in the $\tau_c/\tau_P$ ratio and assumes a Gaussian form for $\tau_c/\tau_P \gtrsim 1$, as depicted in Fig. 4(d). Figure 4(d) also shows that for a given $\tau_c < \tau_P$, the degree of non-Gaussianity increases with the active noise strength, suggesting that rare active bursts are responsible for the exponential side-tails in $P(\Delta x, \tau)$.

In conclusion, we examined the active Brownian ratchet comprising a colloidal particle in an asymmetric periodic potential in the presence of ECP noise that mimics active fluctuations in the living cells. We found that the condition for obtaining maximum velocity and diffusion enhancement is the injection of finite-time-correlated discrete energy pulses rather than the continuous energy supply. This finding is significant for understanding the effect of the non-Gaussian active fluctuation in various physicochemical processes, such as active transport of the molecular motors, inside the living cells. Also, our finding that the non-Gaussian active fluctuation enhances the velocity of Brownian ratchets can be tested in real biological motors.[61-62]

The ECP noise leads to Brownian yet non-Gaussian diffusion. The particle position distribution is always Gaussian, resulting in normal diffusion. In contrast, the distribution of particle displacement can exhibit a non-Gaussian shape with exponential tails when the active force strength is stronger than the thermal strength and each active burst decays fully before the arrival of another burst. Thus, our minimal model is vital in explaining the Brownian yet non-Gaussian diffusion dynamics observed in various active matter systems.[40, 45] This study can be useful in understanding the effect of non-Gaussian active fluctuations on various nonequilibrium and stochastic processes. It can also open new avenues for designing and understanding of efficient synthetic and biological submicron stochastic devices.

ACKNOWLEDGMENTS

This work was supported by the Institute for Basic Science (Grant No. IBS-R020). We thank Juzar Thingna for useful discussions.

# Supplementary Information:

# Transport and Diffusion Enhancement in Experimentally Realized Non-Gaussian Correlated Ratchets

Govind Paneru[1,2], Jin Tae Park[1], and Hyuk Kyu Pak[1,2*]

[1]Center for Soft and Living Matter, Institute for Basic Science (IBS), Ulsan 44919, Republic of Korea
[2]Department of Physics, Ulsan National Institute of Science and Technology, Ulsan 44919, Republic of Korea

E-mail: hyuk.k.pak@gmail.com

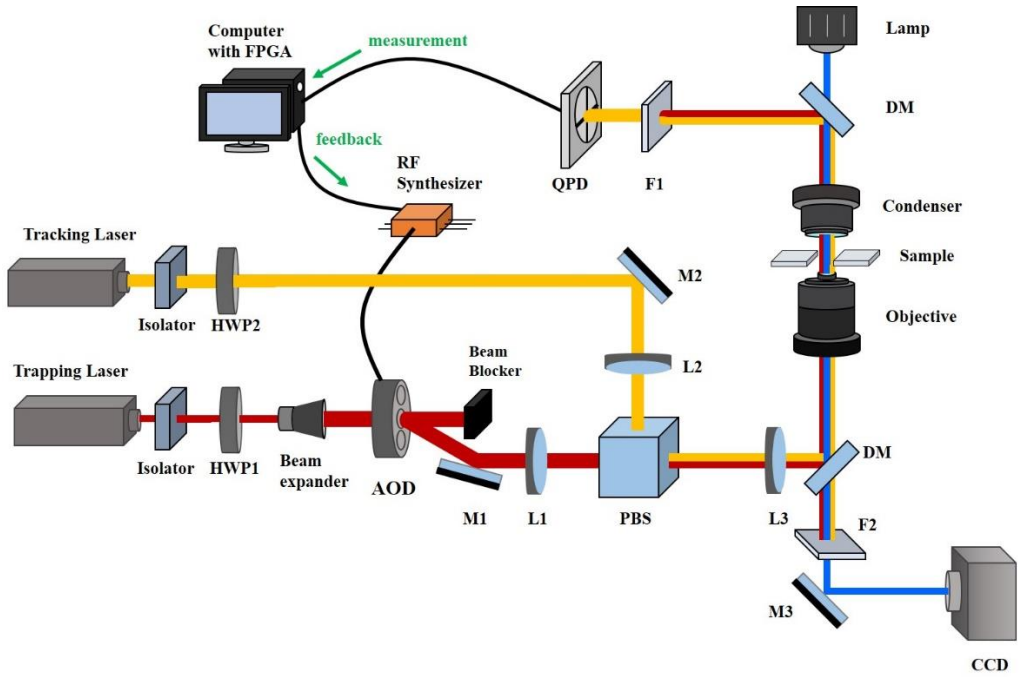

Figure S5. Schematic of the optical tweezers setup. A 1064-nm wavelength laser (trapping laser) is used to trap the particles. The laser beam is passed through an AOD at the Bragg angle, resulting in maximum power output for the first-order beam. This beam is focused on the sample plane of an optical microscope using a 100X oil immersion objective lens with a high numerical aperture (NA = 1.4). A second laser with a wavelength of 980 nm (tracking laser) is used for particle position detection. The particle position is measured by the QPD mounted at the back focal plane of a 1.4 NA condenser lens. The QPD signal is acquired by an FPGA data acquisition board using a custom-written LabVIEW FPGA program. The sample cell includes a highly diluted solution of $2\ \mu$m polystyrene spheres suspended in deionized water at room temperature $297 \pm 1$ K. The trap parameters are calibrated by fitting the probability distribution of the particle position in the thermal equilibrium to the Boltzmann distribution $P(x) = (2\pi\sigma^2)^{-1/2}\exp(-x^2/2\sigma^2)$. The trap stiffness is then estimated as $k \approx 30$ pN$\mu$m$^{-1}$ using the equipartition relation $k = k_B T/\sigma^2$. AOD: acoustic optical deflector, QPD: quadrant photodiode, HWP1, HWP2: half waveplate, M1, M2, M3: mirror. L1, L2, L3: Lens. PBS: polarizing beam splitter, DM: dichroic mirror, F1, F2: filter. CCD: camera.

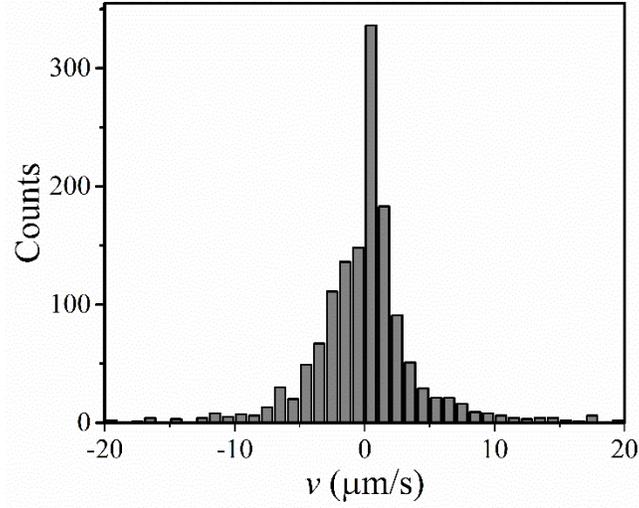

Figure S6. Experimentally measured velocity distribution of the particle in the presence of the exponentially correlated Poisson noise of strength $f_{act} \approx 1.05$ pN with Poisson interval normalized with thermal relaxation time $\tau_P/\tau_r \approx 33.3$ and correlation time $\tau_c/\tau_P \approx 0.25$. The distribution is the result of 1500 individual trajectories. In calculating the mean velocity, we have ignored the trajectories shorter than $10\Delta t$. These trajectories correspond to the initial starting position $x(0)$ near the end of the cutoff distance $\pm 300$ nm, or due to the rare large-amplitude active bursts that kick the particle by several periods of the periodic potential to surpass $\pm 300$ nm within $10\Delta t$.

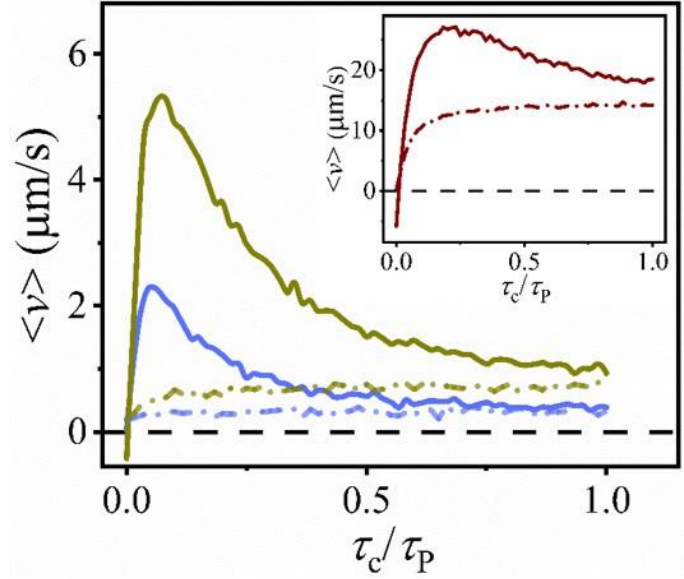

Figure S7. (Numerical Result) Plot of $<v>$ as a function of $\tau_c/\tau_P$ for a $0.5\ \mu$m diameter particle in asymmetric potential of period $L = 8$ nm and barrier height $2V_0 \approx 10 k_B T$ in the presence of ECP noise of fixed $\tau_P/\tau_r = 75$ and noise strength $f_{act} \approx 1.5$ pN (blue solid curve) and 2 pN (dark yellow solid curve). The like color dash dot curves are the plot of $<v>$ in presence of AOU noise as a function of $\tau_c$ normalized with $\tau_P$ of the corresponding ECP noise of the same strength. Inset: Plot of $<v>$ as a function of $\tau_c/\tau_P$ for ECP noise (wine solid curve) and AOU noise (dash dot wine curve) of $f_{act} \sim f_{th} \approx 4$ pN.

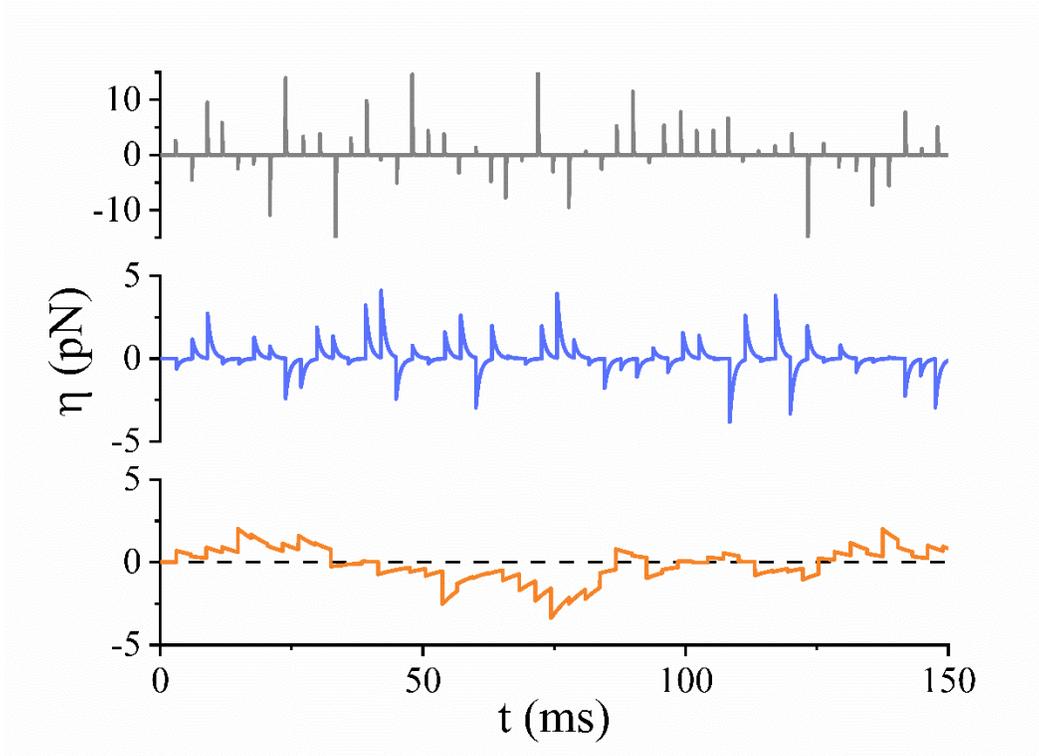

Figure S8. Traces of the ECP noise of strength $f_{act} \sim f_{th} = 0.8$ pN with Poisson interval $\tau_P/\tau_r = 25$ and correlation time $\tau_c/\tau_P = 0.025$ (gray), 0.25 (blue), and 2.5 (orange).